\documentclass[10pt, twocolumn,showpacs,preprintnumbers,amsmath,amssymb]{revtex4-1}

\usepackage{graphicx}% Include figure files
\usepackage{dcolumn}% Align table columns on decimal point
\usepackage{bm}% bold math
\usepackage{epsfig}
\usepackage{graphics}

\begin{document}

\title{\LARGE \bf Control of Inhomogeneous Ensembles on The Bloch Sphere}
%Control of Inhomogeneous Ensembles on the Bloch Sphere 
%Alternatively: Novel Elements for Correcting Pulse Errors

\author{Philip Owrutsky}
\email{philip.owrutsky@gmail.com}
\author{Navin Khaneja}
\email{navin@eecs.harvard.edu}
\homepage{http://hrl.harvard.edu/~navin}
\affiliation{School of Engineering \& Applied Sciences,\\ Harvard University,
Cambridge, MA 02138}
\thanks{The work was supported by ONR 38A-1077404, AFOSR FA9550-05-1-0443 and
AFOSR FA9550-04-1-0427}

\date{\today}

\def\ep {\epsilon}

%%%%%%%%%%%%%%%%%%%%%%%%%%%%%%%%%%%%%%%%%%%%%%%%%%%%%%%%%%%%%%%%%%%%%%%%%%%%%%%%

\begin{abstract}
Finding control fields (pulse sequences) that can compensate for the dispersion in the
parameters governing the evolution of a quantum system is an important problem in coherent
spectroscopy and quantum information processing. The use of composite pulses for compensating
for dispersion in system dynamics is widely known and applied. In this paper, we introduce
new pulse elements for correcting pulse errors. These design methods are analytical and can be
used to prove arbitrarily good robust performance. Furthermore, the time-to-compensation is superior to existing Fourier Synthesis Methods which is critical for minimizing errors due to relaxation effects.  %The increase in the length of the pulse sequence is only polynomial in the level of compensation desired. 
\end{abstract}

%\pacs{03.67.-a}

\maketitle

%%%%%%%%%%%%%%%%%%%%%%%%%%%%%%%%%%%%%%%%%%%%%%%%%%%%%%%%%%%%%%%%%%%%%%%%%%%%%%%%
\section{INTRODUCTION}
Many applications in control of quantum systems involve controlling
a large ensemble by using the same control field. In practice, the
elements of the ensemble could show variation in the parameters
that govern the dynamics of the system. For example, in
magnetic resonance experiments, the spins of an ensemble may have
large dispersion in their natural frequencies (Larmor dispersion),
strength of applied rf-field (rf-inhomogeneity) and the relaxation rates of
the spins \cite{levitt} to name a few. In solid state NMR spectroscopy of powders, the random distribution of
orientations of inter-nuclear vectors of coupled spins within an
ensemble leads to a distribution of coupling strengths \cite{Rohr}. A
canonical problem in control of quantum ensembles
is to develop external excitations that can simultaneously steer the ensemble of
systems with variation in their internal parameters from an initial
state to a desired final state. These are called compensating pulse
sequences as they can compensate for the dispersion in the system dynamics. From the standpoint of mathematical
control theory, the challenge is to simultaneously steer a continuum of
systems between points of interest with the same control signal.
Typical applications are the design of excitation and inversion pulses in NMR
spectroscopy in the presence of larmor dispersion and rf-inhomogeneity
\cite{levitt, tyco, tyco1, shaka, levitt1, levitt2, Garwood, Skinner1, Kobzar1, Pattern}
or the transfer of coherence or polarization in coupled spin ensemble with
variations in the coupling strengths \cite{chingas}. In many cases of
practical interest, one wants to find a control field that prepares the
final state as some desired function of the parameter. For example, slice
selective excitation and inversion pulses in magnetic resonance imaging
\cite{Silver, Rourke, Shinnar, Roux}.
The problem of designing excitations that can compensate for dispersion in the dynamics is
a well studied subject in NMR spectroscopy and
extensive literature exists on the subject of composite pulses
that correct for dispersion in system dynamics \cite{levitt, tyco, tyco1, shaka, levitt1, levitt2, Garwood}.
Composite pulses have recently been used in quantum information processing to correct for
systematic errors in single and two qubit operations \cite{wimperis, jones, brown, molmer, riebe, barrett}.

The focus of this paper is to present novel pulse elements for compensating errors arising from
uncertainties/imperfections in the pulse amplitude. The constructions presented here have the advantage that they are 
analytical and exhibit favorable performance compared to the existing analytical Fourier Synthesis Method \cite{Fourier}.  Namely, a higher level of rf robustness is obtained for the same pulse length and power, which is critical for minimizing errors due to relaxation.   %Originally ellipses

To fix ideas, consider an ensemble of noninteracting spin $\frac{1}{2}$ particles in a
static field $B_0$ along the $z$ axis and a transverse rf-field,
$(A(t)\cos(\phi(t)), A(t) \sin(\phi(t)))$, in the $x-y$ plane.
Let $x, y, z$ represent the coordinates of the unit vector in the direction of the net magnetization vector of the ensemble.
The dispersion in the amplitude of the rf-field is given by a dispersion parameter $\epsilon$ such that
$A(t) = \epsilon A_0(t)$ where $\epsilon \in [1 - \delta, 1 + \delta]$, for $\delta > 0$.  The Bloch equations for the ensemble
take the form
\begin{equation}
\label{eq:Bloch1}
\frac{d}{dt}{\left[\begin{array}{c}x\\y\\z\end{array}\right]}
=\left[\begin{array}{ccc}0&-\omega& \epsilon u(t)\\
\omega&0& -\epsilon v(t)\\-\epsilon u(t)& \epsilon v(t)&0\end{array}\right]
\left[\begin{array}{c} x\\y\\z\end{array}\right],
\end{equation}where $$(u(t), v(t))= \gamma (A_0(t)\cos(\phi(t)), A_0(t) \sin(\phi(t))). $$
Let $X(t)$ denote the unit vector $(x(t), y(t), z(t))$.
Consider now the problem of designing controls $u(t)$ and $v(t)$
that simultaneously steer an ensemble of such systems with dispersion in the strength of their  rf-field from an initial state
$X(0)=(0,0,1)$ to a final state $X_F=(1,0,0)$
\cite{Skinner1}. This problem raises interesting questions about
controllability, i.e., showing that in spite of bounds on the strength of the rf-field,
$\sqrt{u^2(t)+v^2(t)}\leq A_{max}$, there exist excitations $(u(t), v(t))$,
which simultaneously steer all the systems with dispersion in
 $\ep$, to a ball of desired radius $r$ around the
final state $(1,0,0)$ in a finite time (which may depend on $A_{max}$, $B$, $\delta$, and $r$).
Besides steering the ensemble between two points, we can ask for a control field that steers an
initial distribution of the ensemble
to a final distribution, i.e., different elements of the ensemble now have different initial and final states
depending on the value of the their dispersion parameter $\ep$. The initial and final state of the ensemble
are described by functions $X_0(\ep)$ and $X_F(\ep)$ respectively. Consider the problem of steering an
initial distribution $X_0(\ep)$ to within a desired distance $r$ of a target function $X_F(\ep)$ by
appropriate choice of controls in equation (\ref{eq:Bloch1}).  We use the L2 norm as our error metric
\begin{eqnarray}
E^2 = \int_{1-\delta}^{1+\delta} ||X_F (\epsilon)-X_{target}(\epsilon)||^2 d\epsilon.
\end{eqnarray}
If a system with dispersion in its parameters can be steered between states that have dependency on the
dispersion parameter arbitrarily well, then we say that the system is ensemble controllable with respect
to those parameters.

This paper is organized as follows. In the following section, we introduce the key ideas and through examples, highlight the role of non-commutativity in the design of a compensating control. We present novel pulse elements that compensate for inhomogeneity or uncertainty in the amplitude of the rf-field.  The presented method extends known techniques for pulse sequences that are robust to rf inhomogeneity.  The methods presented may also find applications in design of anisotropy compensating pulse design or in solid state NMR experiments \cite{Comp_Dipolar}.

%%%%%%%%%%%%%%%%%%%%%%%%%%%%%%%%%%%%%%%%%%%%%%%%%%%%%%%%%%%%%%%%%%%%%%%%%%%%%%%%
\section{Novel Pulse Elements for Compensating Rf-Inhomogeneity}

\subsection{Fourier Synthesis Technique}

\label{sec:Lie.Bracket}

\noindent \textbf{Example 1: Main Concept} \\ \noindent {\rm To fix ideas, we begin by considering the Bloch
equations in a rotating frame with only
rf-inhomogeneity and no Larmor dispersion.
\begin{equation}
\label{eq:main.1}
\dot{X} = \epsilon (u(t) \Omega_y + v(t) \Omega_x )X,
\end{equation}where
$$\Omega_x = \left[\begin{array}{ccc}0& 0 & 0 \\
0 & 0& -1 \\ 0 & 1 & 0 \end{array}\right],\ \ \Omega_y = \left[\begin{array}{ccc}0& 0 & 1 \\
0 & 0& 0 \\ -1 & 0 & 0 \end{array}\right], \Omega_z = \left[\begin{array}{ccc}0& -1 & 0 \\
1 & 0& 0 \\ 0 & 0 & 0 \end{array}\right] $$
are the generators of rotation around
$x$, $y$ and $z$ axis, respectively. \\

\noindent We define the pulse elements \cite{Fourier}
\begin{eqnarray}
\label{eq:sequencepulse}
U_1(\beta_k) &=&  \exp(-\gamma_k  \Omega_x \epsilon ) \exp(\frac{\beta_k}{2} \Omega_y \epsilon )\exp( \gamma_k  \Omega_x \epsilon), \\
U_2(\beta_k) &=&  \exp(\gamma_k  \Omega_x \epsilon ) \exp(\frac{\beta_k}{2} \Omega_y \epsilon)\exp(- \gamma_k  \Omega_x \epsilon).
\end{eqnarray}
which correspond to directly accessible evolutions.  

For suitably small $\beta_k$ , we have
\begin{eqnarray}
 V_k = U_2 U_1 \sim \exp(\epsilon \beta_k \Omega_y \cos(\gamma_k  \epsilon)) \label{eq:vk}. 
 \end{eqnarray}
To effect a larger amplitude rotation, we consider the sequence of transformations
\begin{equation}
\label{eq:compositepulse}
H_1 \equiv \prod_{k} (V_k)^{n_k} \sim \exp \left(\epsilon \sum_k \alpha_k \cos(\gamma_k  \epsilon) \Omega_y \right),
\end{equation}
where $\alpha_k = n \beta_k$.  In practice, $\beta_k < \frac{\pi}{10}$ is suitably small and results in an error that is less than 1\% in the L2 sense in (\ref{eq:vk}).   Now, the coefficients $\alpha_k$ can be chosen so that 
\begin{eqnarray}
\sum_k \alpha_k \cos(\gamma_k  \epsilon) \approx \frac{\theta}{\epsilon} \label{eq:expansion1}
\end{eqnarray}
for $1 - \delta \leq \epsilon \leq 1+\delta$, with $0 < \delta < 1$.

\noindent Therefore,
\begin{equation}
\label{eq:net_FSM} 
H_1 \sim \exp(\theta \Omega_y) %\prod_{k} (V_k)^{n_k}
\end{equation}
approximately independent of $\epsilon$.  The dependence on $\epsilon$ can be made arbitrarily small by increasing the pulse length and extending the number of terms in the summation leading to pulses that are immune to dispersions in the rf amplitude as claimed.

\subsection{Modified Fourier Synthesis Using $\delta$ Modulation}

%2 term Expansion Gradient Descent
\begin{figure*}[!t]
\begin{minipage}[b]{.95\columnwidth}
\centering
\includegraphics[width=\columnwidth,trim=40 180 60 180,clip=]{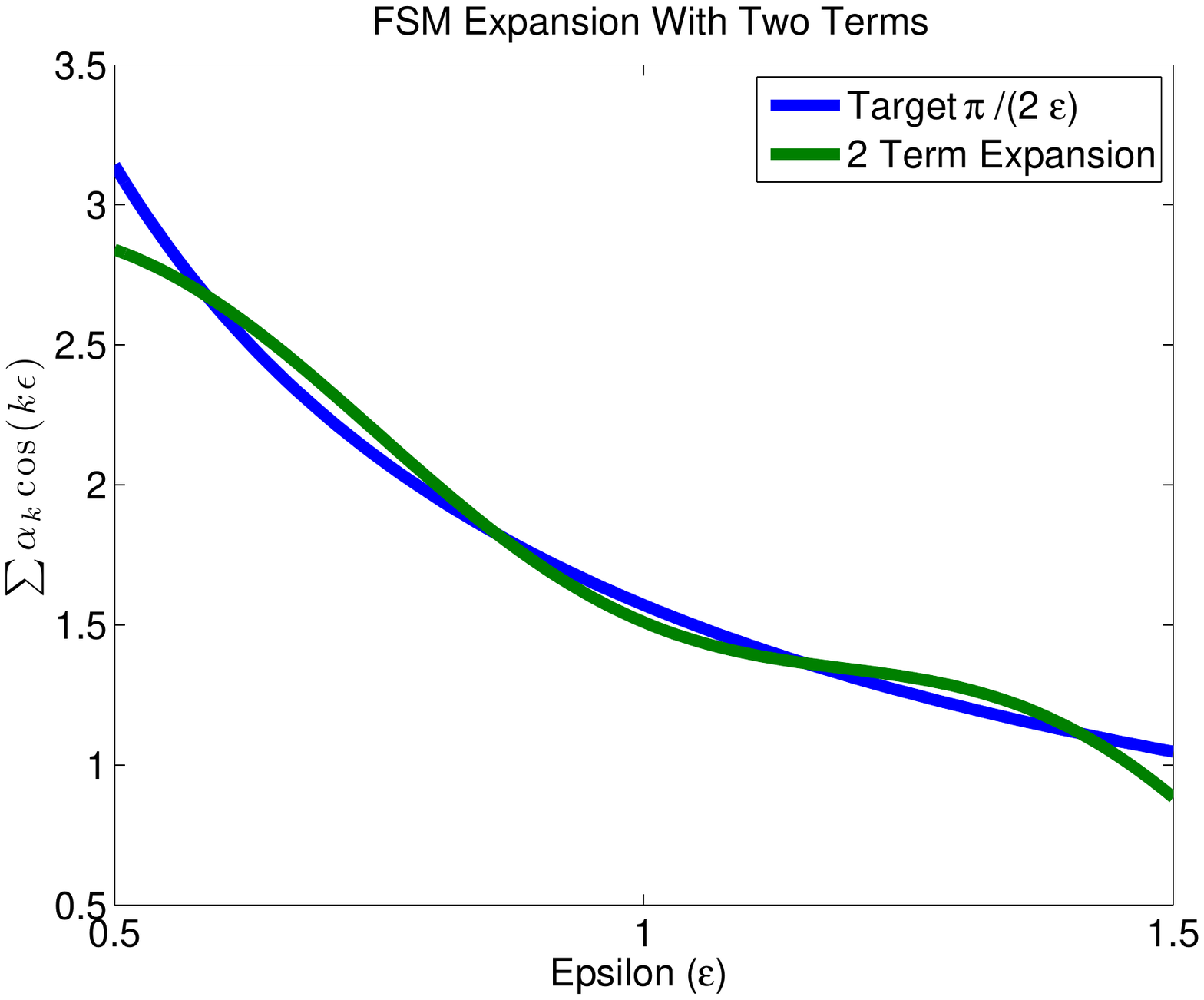}
\end{minipage} \hspace{1cm}
\begin{minipage}[b]{.95\columnwidth}
\centering
\includegraphics[width=\columnwidth,trim=40 180 60 180,clip=]{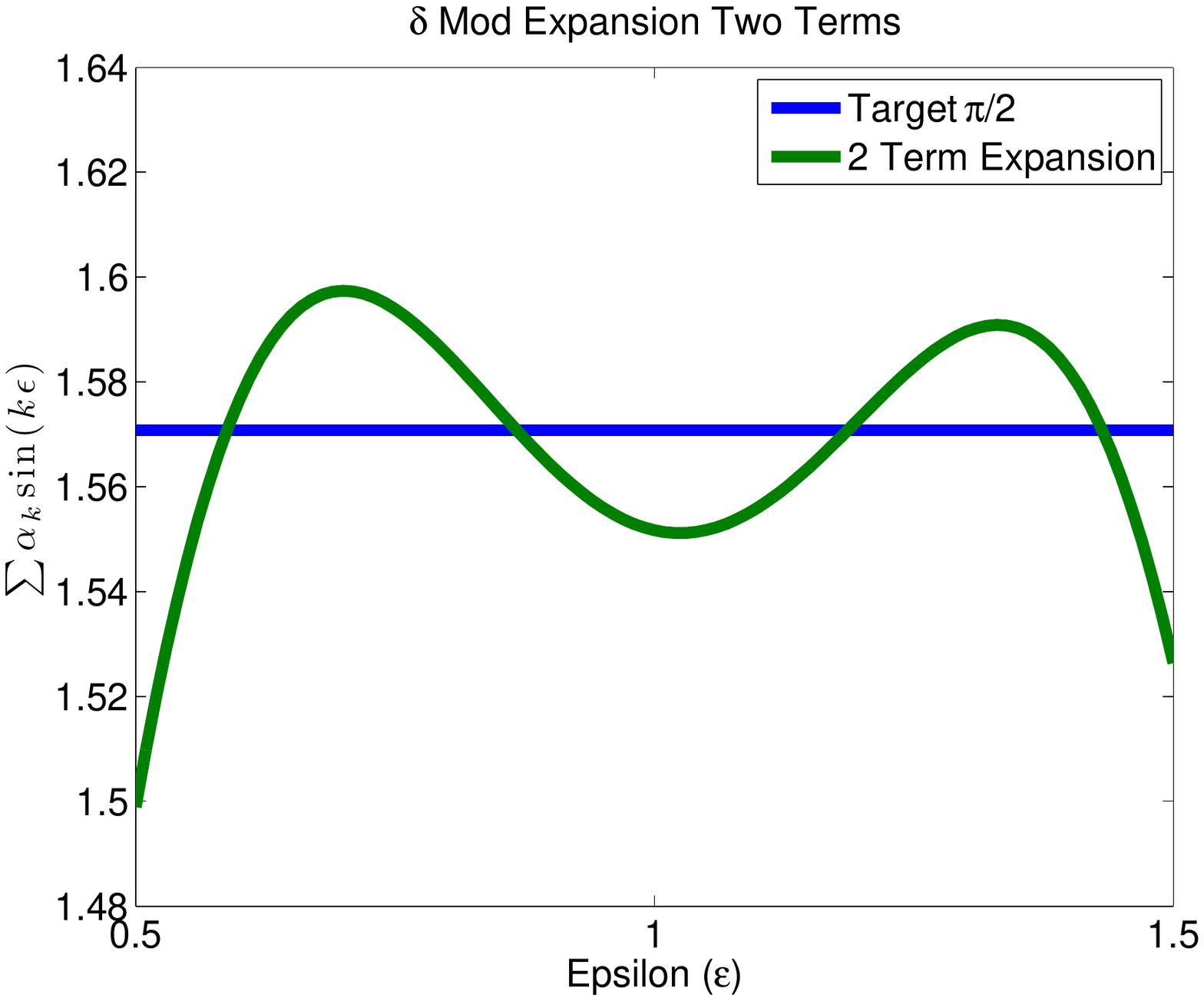}
\end{minipage}
\caption{Left: 2 Term approximation for $H_1=\frac{\pi}{2\ep}$.  Right: 2 Term approximation for $H_2=\frac{\pi}{2}$ using two terms with gradient descent for frequency selection}
\end{figure*}

In this section we develop a modified Fourier Synthesis Method that will be shown to have favorable time-robustness properties to the original Fourier Synthesis technique.  To this end we consider the following system
\begin{equation}
\label{eq:bloch_y}
\dot{Y} = (\epsilon u(t)\Omega_x + v(t)\Omega_z)Y
\end{equation}
which corresponds to a system with one pure control by way of $v(t)$ and one control with dispersion, $\epsilon u(t)$.  We will apply a similar Fourier Synthesis Method (FSM) analysis on the system and show that this results in a modified Hamiltonian to that of the previous section, the advantages of which will be discussed in section III.  We then show how the previous system with both controls exhibiting dispersion (\ref{eq:main.1}) can be transformed into (\ref{eq:bloch_y}) by an appropriate  change of coordinates.

%We now apply Fourier Synthesis Methods (FSM) under the assumption that we have a pure control and show this similarly compensates for the dispersion.  Then we show how this modified FSM can be directly implemented using the available corrupted controls. These modified FSM pulses exhibit favorable compensation for duration characteristics than standard FSM making them of practical interest.  

Consider the modified transformations
\begin{eqnarray}
\tilde{U}_1 &=&  \exp(-\gamma_k \epsilon \Omega_x) \exp(\frac{\beta_k}{2} \Omega_z)\exp(\gamma_k  \epsilon \Omega_x) \\
 \tilde{U}_2 &=&  \exp(\gamma_k  \epsilon \Omega_x) \exp(-\frac{\beta_k}{2} \Omega_z)\exp(-\gamma_k  \epsilon \Omega_x)    
\end{eqnarray}
By again choosing $\beta_k$ sufficiently small, we have
\begin{eqnarray} \label{eq:delta_V}
\tilde{V}_k = \tilde{U}_2 \tilde{U}_1 \sim \exp( \beta_k \sin(\gamma_k \epsilon) \Omega_y). 
\end{eqnarray}
Applying a sequence of such transformations
\begin{equation}
\label{eq:compositepulse}
H_2 \equiv \prod_{k} (\tilde{V}_k)^{n_k} \sim \exp \left( \sum_k \alpha_k \sin(\gamma_k \epsilon) \Omega_y \right),
\end{equation}
where again $\alpha_k = n \beta_k$ is used to control the error from the approximation in (\ref{eq:delta_V}). Now, the coefficients $\alpha_k$ and $\gamma_k$ can be chosen so that 
\begin{equation}
\label{eq:net_MFSM}
 \sum_k \alpha_k \sin(\gamma_k \epsilon) \approx \theta,
\end{equation}
over the range of $\epsilon$ of interest $1 - \delta \leq \epsilon \leq 1 + \delta$ resulting in a robust rotation.  We point out that (\ref{eq:net_MFSM}) resembles (\ref{eq:expansion1}), but no longer contains an $\epsilon$ factor external to the trigonometric argument and that $\cos$ has been replaced with $\sin$.

To see how (\ref{eq:bloch_y}) can be generated from (\ref{eq:main.1}), we return to (\ref{eq:main.1}) 
\begin{eqnarray*}
\dot{X} = \epsilon A ( \cos \underbrace{(\phi_1(t) + \phi_2(t))}_{\phi(t)} \Omega_x + \sin \underbrace{(\phi_1(t) + \phi_2(t))}_{\phi(t)} \Omega_y )X  %(\ref{eq:main.1}b)
\end{eqnarray*}
and move into the frame
\begin{eqnarray}
Y &=& \exp(-\phi_2(t) \Omega_z) X \\
 \dot{Y} &=& [ \epsilon A(t) ( \ \cos \phi_1(t) \Omega_x +  \sin \phi_1(t)\Omega_y \ ) - \dot{\phi_2} \Omega_z ] Y 
 % \dot{Y} &=& [ \epsilon \tilde{u}(t) \Omega_x + \epsilon \tilde{v}(t) \phi_1(t)\Omega_y \ ) - \dot{\phi_2} \Omega_z ] Y
 \end{eqnarray}
which corresponds to (\ref{eq:bloch_y}) once the appropriate identifications are made.  

This means that $\tilde{V}_k$ can be directly produced by implementing $\phi_1(t)$ as $0$, $\pi$, $\pi$ and $0$ over $\Delta t$ time intervals such that $A \Delta t = \gamma_k $, and with $-\dot{\phi}_2$ a delta pulse with area $\frac{\alpha_k}{2}$, and $-\frac{\alpha_k}{2}$ at time $\Delta t$ and $3 \Delta t$ respectively in (\ref{eq:main.1}). As $\phi_2(4\Delta t)=0$, $X(4\Delta t)=Y(4\Delta t)$ and the lab frame and Y frame coincide after each pulse sequence, completing the $\delta$ Modulated Pulse Design Method.

%We complete the design algorithm by showing $\tilde{V}_k$ can be directly implemented as
%\begin{equation}
%\tilde{V}_k = \exp(k \theta \epsilon \Omega_x) \exp(2 k \theta \epsilon \Omega_{\phi})\exp(k \theta \epsilon \Omega_x)
%\end{equation}
%
%\noindent with $\Omega_{\phi} = \Omega_x \cos \phi + \Omega_y \sin \phi $ and $\phi = 180 - \frac{\beta_k}{2}$.  This can be seen by returning to (1) as follows
%\begin{equation}
%\dot{X} = \epsilon A ( \cos \underbrace{(\phi_1(t) + \phi_2(t))}_{\phi(t)} \Omega_x + \sin \underbrace{(\phi_1(t) + \phi_2(t))}_{\phi(t)} \Omega_y )X,
%\end{equation}
%now transform into the frame 
%\begin{eqnarray*}
% Y &=& \exp(-\phi_2(t) \Omega_z) X \\
% \dot{Y} &=& ( \epsilon A (  \cos \phi_1(t) \Omega_x +  \sin \phi_1(t)\Omega_y) - \dot{\phi_2} \Omega_z ) Y
% \end{eqnarray*}
%Choosing $\phi_1(t)$ as $0$, $\pi$, $\pi$ and $0$ over $\Delta t$ time intervals each
%such that $A \Delta t = k \theta$, and with $-\dot{\phi}_2$ a delta pulse with area $\frac{\alpha_k}{2}$, and
%$-\frac{\alpha_k}{2}$ at time $\Delta t$ and $3 \Delta t$ respectively, then to first order
%\begin{equation}
%Y(4 \Delta t)  = \tilde{U}_2\tilde{U}_1 ,
%\end{equation}
%The effect of the delta pulse $-\dot{\phi}_2$ is a discontinuous change in the phase $\phi(t)$ and therefore
%\begin{equation}
%X(4 \Delta t) = \exp(k \theta \epsilon \Omega_x) \exp(2 k \theta \epsilon \Omega_{\phi})\exp(k \theta \epsilon \Omega_x)
%\end{equation}
%Since $\phi_1(4 \Delta_t) = 0$, $X_4(\Delta t) = Y_4(\Delta t)$ showing $(3) \equiv (5)$ and completing the pulse design algorithm.  

The only remaining task is selection of the amplitude and frequency coefficients which we defer until the following section where we discuss the associated performance of the algorithms.

In this paper we have focused on producing $\epsilon$-robust uniform rotations, however the reader should note that by appropriate selection of $\alpha_k$ in $H_1$ and $H_2$, any rotation as a function of $\epsilon$ can be produced. \vspace{.7cm}

%%%%%%%%%%%%%%%%%%%%%%%%%%%%%%%%%%%%%%%%%%%%%%%%%%%%%%%%%%%%%%%%%%%%%%%%%%%%%%%%

\section{Simulations and Error Performance of F. Synthesis and $\delta$ Modulation}

%2 term expansion Error Gradient Descent
\begin{figure*}[!t]
\begin{minipage}[b]{.95\columnwidth}
\centering
\includegraphics[width=\columnwidth,trim=40 180 60 180,clip=]{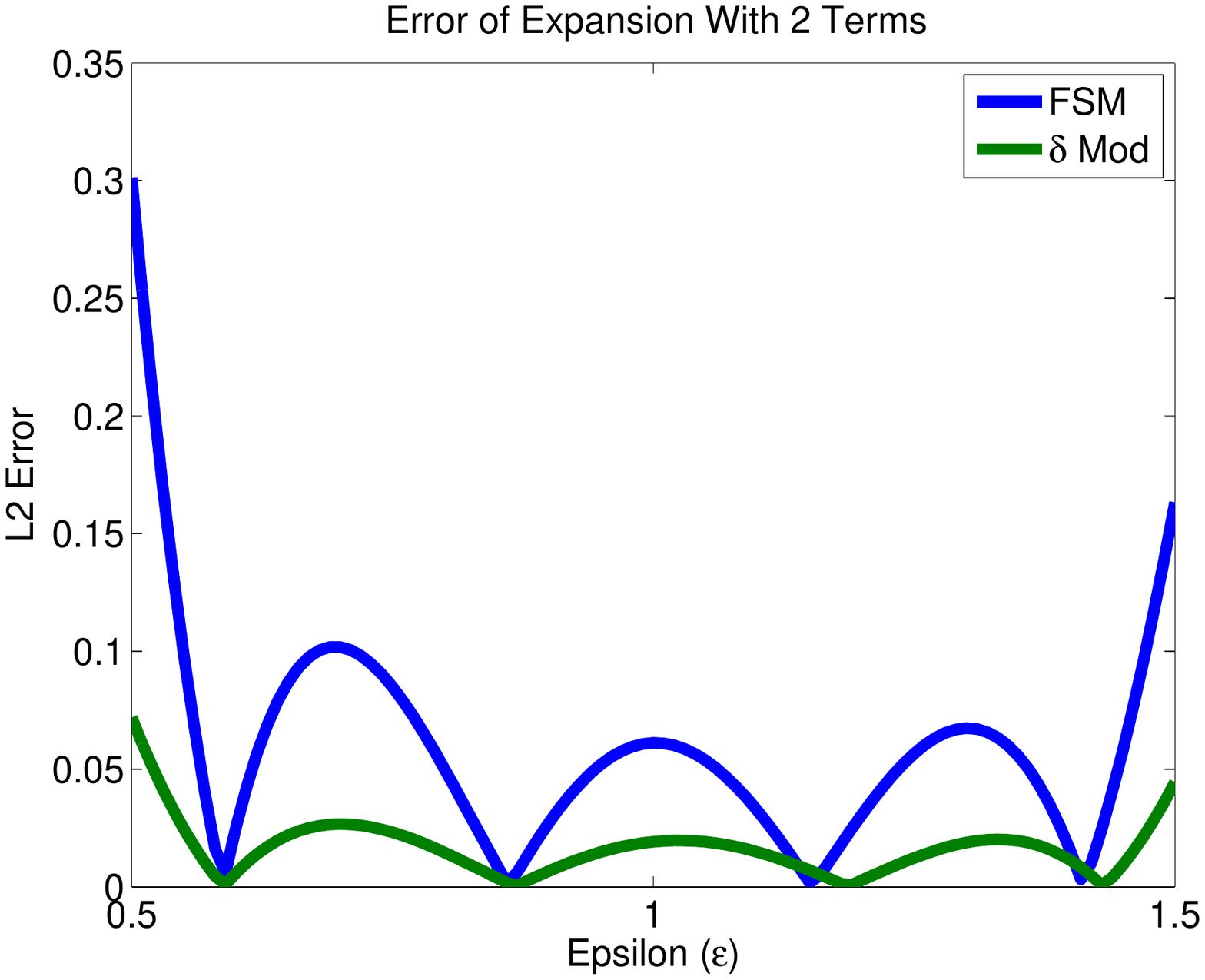}
\end{minipage} \hspace{1cm}
\begin{minipage}[b]{.95\columnwidth}
\centering
\includegraphics[width=\linewidth,trim=30 180 60 180,clip=]{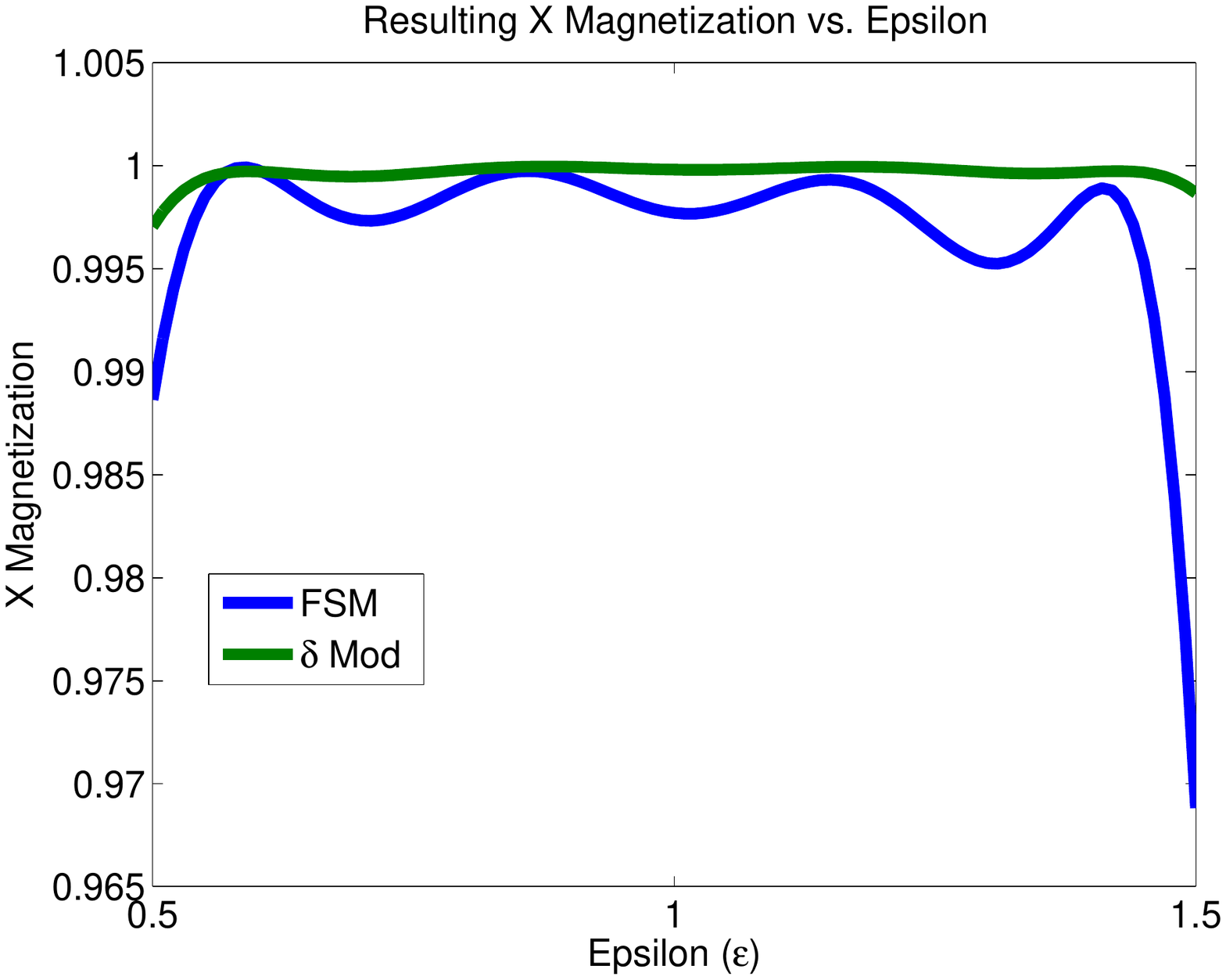}
\end{minipage}
\caption{Left:  L2 Error with respect to the desired final magnetization $[1,0,0]'$ for two term $H_1$ and $H_2$ as a function of the dispersion parameter $\ep$ where frequencies were found using gradient descent.  Right:  Resulting X-magnetization for the 2 term sequence for both Fourier Synthesis and $\delta$ Modulation.  $\delta$ modulation has a favorable magnetization profile while requiring a shorter pulse duration. }
\end{figure*}

%%%%%%%
\begin{widetext}{
%\begin{table}[]
\begin{minipage}[b]{.29\columnwidth}
\centering
%\begin{table}
\begin{tabular}{lccc}
\textbf{Heuristic} \\
& n=2 & n=3 & n=4 \\
Error FSM & 0.06831 & 0.06523 & 0.06473  \\
Error $\delta$ Mod & 0.02012 & 0.00290 & 0.00044 \\
Flip $\angle$ FSM &  127.120 & 187.200 & 199.600 \\
Flip $\angle$ $\delta$ Mod & 115.230 & 172.001 & 216.138
\end{tabular}
%\end{table}
\end{minipage} \hspace{1cm}
\begin{minipage}[b]{.29\columnwidth}
\centering
%\begin{table}
%\caption{Greedy Algorithm Frequency}
\begin{tabular}{lccc}
\textbf{Greedy}\\
& n=2 & n=3 & n=4 \\
Error FSM & 0.04031 & 0.01506 & 0.00941  \\
Error $\delta$ Mod & 0.02029 & 0.00422 & 0.00247 \\
Flip $\angle$ FSM &  130.497 & 179.062 & 229.101 \\
Flip $\angle$ $\delta$ Mod & 110.952 & 165.178 & 216.177
\end{tabular}
%\end{table}
\end{minipage} \hspace{1cm}
\begin{minipage}[b]{.29\columnwidth}
\centering
%\begin{table}
%\caption{Gradient Descent Frequency}
\begin{tabular}{lccc}
\textbf{Gradient} \\
& n=2 & n=3 & n=4 \\
Error FSM & 0.07339 & 0.01874 & 0.00423 \\
Error $\delta$ Mod & 0.01940 & 0.00280 & 0.00044 \\
Flip $\angle$ FSM &  120.519 & 225.780 & 347.413  \\
Flip $\angle$ $\delta$ Mod & 113.341 & 170.134 & 216.137
\end{tabular}
\end{minipage} }
%\end{table}
\\ Table 1:    Performance of Fourier Synthesis and $\delta$ modulation for heuristic, greedy and gradient descent based frequency selection.  In all cases, $\delta$ modulation outperforms the Fourier Synthesis method in terms of L2 Error for a given pulse duration, which is important for minimizing relaxation effects.  L2 error is  calculated with respect to the desired final magnetization $[1,0,0]'$.
%\end{table} 
\end{widetext}
%%%%%%

The previous section reduced the problem of RF dispersion compensation to parameter selection, $\gamma_k$ and $\alpha_k$.  Given the inhomogeneity parameter, $\epsilon \in [1- \delta_0, 1 + \delta_0] $, we compute the error performance of synthesizing the effective Hamiltonians
\begin{eqnarray}
H_1 &=& \exp \left( \epsilon \sum_k \alpha^1_k \cos(\epsilon \gamma^1_k ) \Omega_y \right) \\ 
H_2 &=& \exp \left( \sum_k \alpha^2_k \sin(\epsilon \gamma^2_k) \Omega_y \right)
\end{eqnarray} to approximate
\begin{eqnarray*}
H = \exp(\theta \Omega_y).
\end{eqnarray*}
We note that the $\{\alpha^1_k\}$ can be directly calculated given the $\{\gamma^1_k\}$ as
 \begin{eqnarray}
 \vec{\alpha}^1 = \Phi^{-1}V; \quad \langle f,g  \rangle = \int_{1-\delta}^{1+\delta} f(\ep)g(\ep)d\ep  \label{eq:alpha1} \\
 \Phi_{ij} = \langle \cos(\gamma^1_i \ep) ,\cos(\gamma^1_j \ep)  \rangle; \quad V_i = \langle \cos(\gamma^1_i \ep), \frac{\theta}{\ep} \rangle
 \end{eqnarray}
and similarly the $\{\alpha^2_k\}$ can be calculated as
\begin{eqnarray}
 \vec{\alpha}^2 = \Phi^{-1}V; \quad \langle f,g  \rangle = \int_{1-\delta}^{1+\delta} f(\ep)g(\ep)d\ep \\
 \Phi_{ij} = \langle \sin(\gamma^2_i \ep) ,\sin(\gamma^2_j \ep)  \rangle; \quad V_i = \langle \sin(\gamma^2_i \ep), \theta \rangle \label{eq:alpha2}
 \end{eqnarray}
so that the problem reduces to selecting the optimal frequencies.  

We report performance for three frequency selection methods, heuristically, greedy selection and gradient descent and show that $\delta$ modulation outperforms Fourier Synthesis Methods for all frequency selection methods. Unless stated otherwise, the notion of optimal is with respect to L2 error for a given pulse duration.  L2 error is calculated with respect to the desired final magnetization $[1,0,0]$ and pulse duration is reported in total flip angle. 

As a starting point, we consider the problem of selecting the optimal amplitudes given known frequencies which we will choose heuristically.  As sine obtains its maximum at $\pi/2$ and is relatively horizontal about this point, a natural selection for the frequencies in (\ref{eq:net_MFSM}) is $\gamma_k = \frac{(2k+1)\pi}{2}$.  Similarly, selecting $\gamma_k$ in (8) to maximize flatness  about $\ep=1$ corresponds to
\begin{eqnarray}
\left[ \frac{d}{d\epsilon} \epsilon \cos(\gamma_k \epsilon) \right] \bigg|_{\ep = 1}  = \cos(\gamma_k) - \gamma_k \sin(\gamma_k) = 0.
\end{eqnarray}
Numerically solving gives the first several $\gamma_k=[.860,3.426,6.437]$.  

The amplitude coefficients $\vec{\alpha}$ were then calculated according to (\ref{eq:alpha1})-(\ref{eq:alpha2}).  The comparative performance of standard FSM, $H_1$, to $\delta$ modulation, $H_2$, is tabulated in table 1 and a complete description of the pulses is given in the appendix.

An alternative algorithm is to sequentially select the frequencies employing a greedy algorithm, in which already determined frequencies are held fixed, and only the newest frequency is optimized over.  Explicitly we sequentially minimize the cost functions with respect to $\gamma_k^{1/2}$
\begin{eqnarray}
F_1(\gamma^1_k,...,\gamma^1_1 ) = \int_{1-\delta}^{1+\delta} \left| \left| \sum \alpha_k \cos(\epsilon \gamma^1_k) - \frac{\theta}{\epsilon} \right| \right| d\epsilon \\
F_2 (\gamma^2_k,...,\gamma^2_1 ) = \int_{1-\delta}^{1+\delta} \left| \left| \sum \alpha_k \sin(\epsilon \gamma^2_k) - \theta \right| \right| d\epsilon
\end{eqnarray}
again using (\ref{eq:alpha1})-(\ref{eq:alpha2}) to calculate the $\vec{\alpha}$.  This was done using gradient descent and numerically calculating the necessary derivatives.  Table 1 shows that the $\delta$ modulation outperforms standard FSM.  %As the previous frequencies were a natural choice for $H_2$, the greedy algorithm compares the methods on a more equal footing.  

The most general method we applied (and best performing) was simultaneously optimizing $F_1$ and $F_2$ with respect to all frequencies using gradient descent, where derivatives were again calculated numerically.  As with all descent schemes, there is concern that one merely obtains a local minima.  Moreover, the problem of unspecified frequencies is how to project onto an overrepresented subspace which is known to have local minima.  To combat such issues we chose the optimal result after numerous starting points and note that the performance exceeds the other methods and the results are displayed in table 1.

As an example we consider the resulting parameters from optimizing a two term $\delta$ modulated pulse using gradient descent
\begin{eqnarray*}
\gamma^2 = [88.6^\circ, 265.1^\circ]; \qquad \alpha^2 = [105.5^\circ, 16.6^\circ].
\end{eqnarray*}
These are converted into a pulse sequence by first dividing large amplitudes of $\alpha_k^2$ into repeated sequences with smaller amplitudes according to (14) using a threshold value of $9^\circ$, which yields the modified parameters
\begin{eqnarray*}
\gamma^{2'} &=& [\underbrace{88.6^\circ,...,88.6^\circ}_{\mbox{12 times}}, \underbrace{265.1^\circ,...,265.1^\circ}_{\mbox{2 times}}] \\
\alpha^{2'} &=& [\underbrace{8.8^\circ,...,8.8^\circ}_{\mbox{12 times}}, \underbrace{8.4^\circ,...,8.4^\circ}_{\mbox{2 times}}]  
\end{eqnarray*}
Pulses are calculated as described in section II.B where pulse elements are 
\begin{eqnarray*}
[(\gamma_k)_0(2\gamma_k)_{180^\circ -\alpha_k/2}(\gamma_k)_0]
\end{eqnarray*}
with numbers inside the parentheses representing the flip angle and the subscripts, the phase.  Applying to the modified parameters above yields the pulse sequence
\begin{eqnarray*}
\left[ (88.6)_0(177.1)_{175.6}(88.6)_0 \right]^{\times 12}   \left[ (265.1)_0(530.1)_{175.9}(265.1)_0  \right]^{\times 2}.
%(88.6)_0(177.1)_{105}(88.6)_0(265.1)_0(530.1)_{16.6}(265.1)_0
\end{eqnarray*}
The performance is displayed in figure 2.  The more terms kept in the series, the longer the sequences and overall pulse, but the more robust.  %These are generalizations of existing composite pulses such as the well known $(90)_0(180)_{90}(90)_0$ pulse \cite{levitt}.

%We should detail the computation of the parameters $\beta_k$ and the choice of $theta_k$. 
%The first guess on $theta_k = (2k-1)*\frac{\pi}{2}$. We talked about alternate ways to choose
%the basis $\theta_k$. Error performance for $k=1, 2, ..N$ may be tabulated as a plot
%for both versions. 
%
%\subsection{Pulse Parameters}
%$ (90)_{0}(180)_{178}90_{0}......$
%
%Number inside the parenthesis is the flip angle and subscript is the phase. The more then number
%of terms kept in the series, the longer the sequences. It will be worthwhile explicitly providing some such sequences. 
%
%Generalization of known basic blocks like $(90)_0 180_{90} 90_{0}$. 

\section{General Modulation Schemes}

\begin{figure}[!t]
\centering
\includegraphics[width=\linewidth,trim=20 0 20 0,clip=]{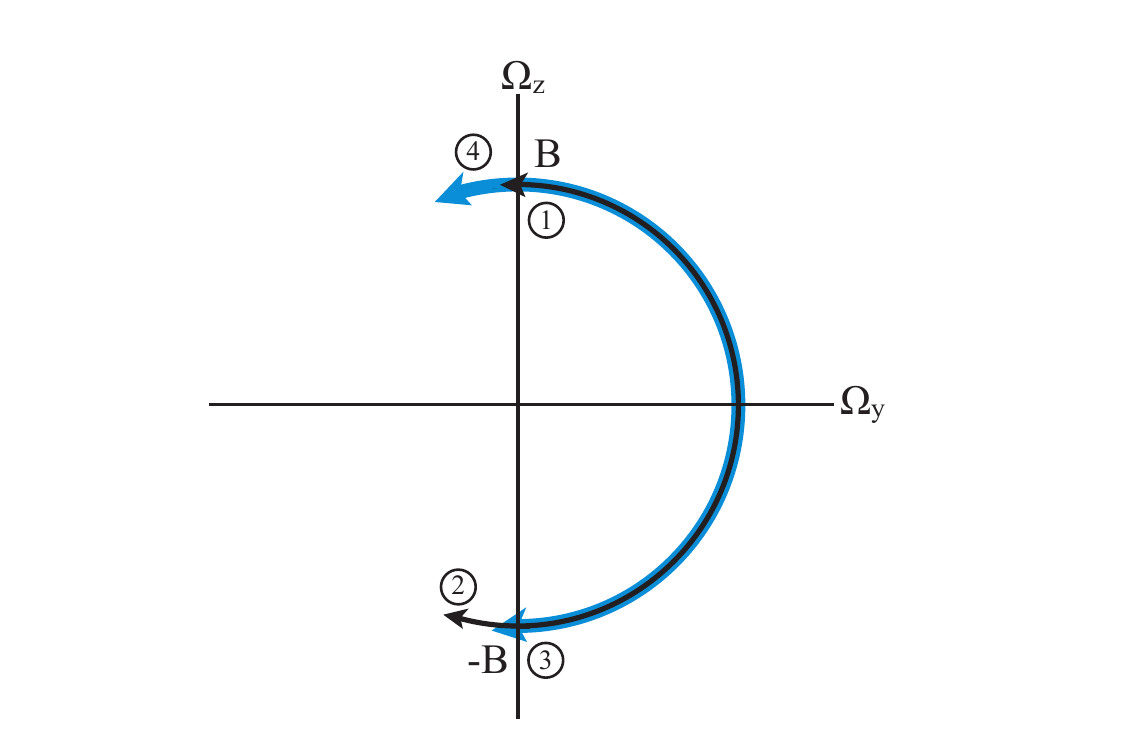}
\caption{Trajectory in the interaction frame for linearly modulated controls.  The trajectory is the black path from 1-2 and back to 1, followed by the blue path from 3-4 and back to 3.}
\end{figure}

In many ways $\delta$ modulation is the most natural choice as it has a nice correspondence with existing FSM's.  However, other modulation schemes are possible and their analysis is warranted for the sake of completeness or in the event that abrupt phase adjustments in the RF-fields are not available.  We begin by considering linear modulation.

\subsection*{Linear Modulation}

%%%%%%%%%%%%%%%%%%%%
%Plots

\begin{figure*}[t!]
\begin{minipage}[b]{.95\columnwidth}
\centering
\includegraphics[width=\columnwidth,trim=40 180 60 180,clip=]{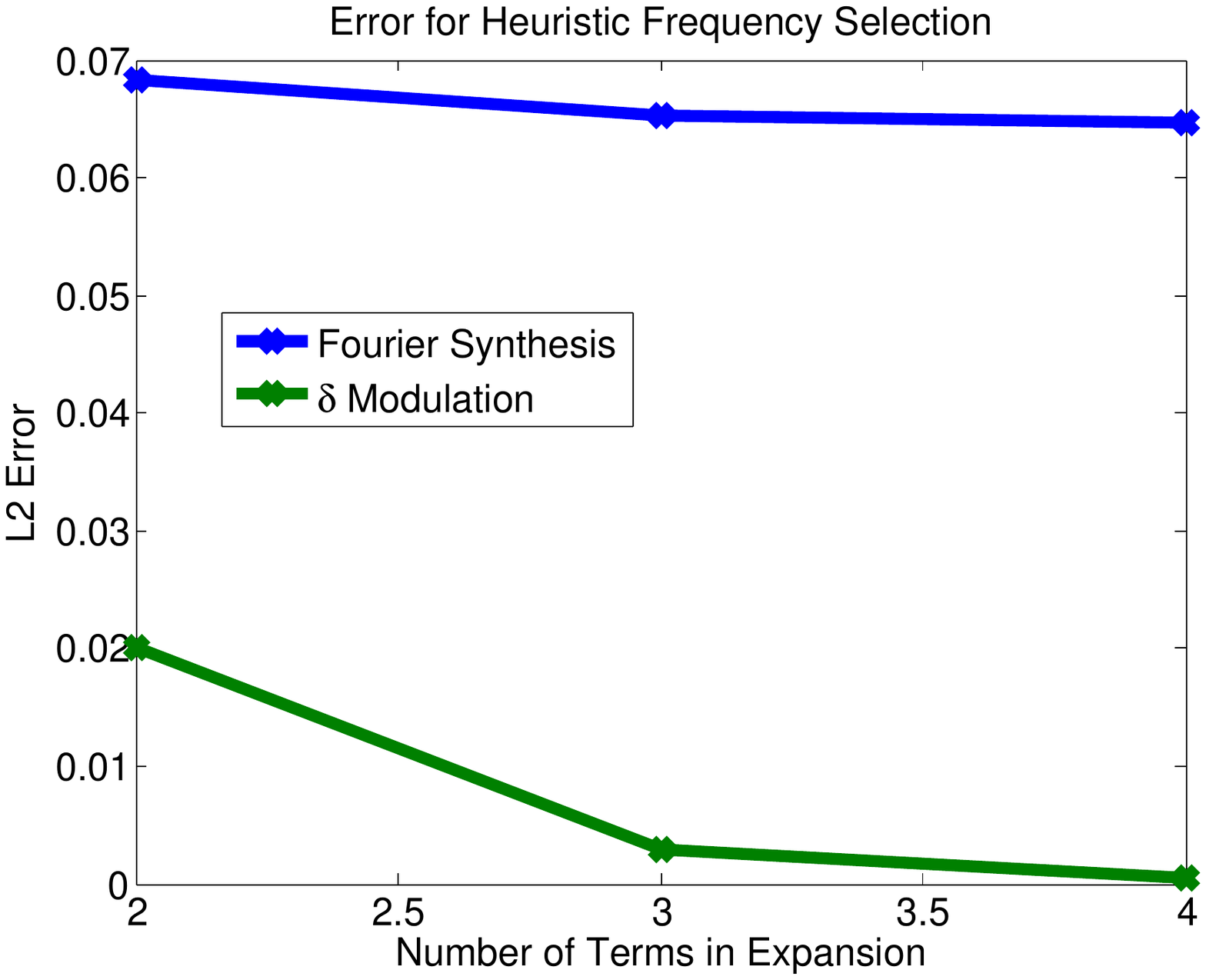}
\end{minipage} \hspace{1cm}
\begin{minipage}[b]{.95\columnwidth}
\centering
\includegraphics[width=\linewidth,trim=30 180 60 180,clip=]{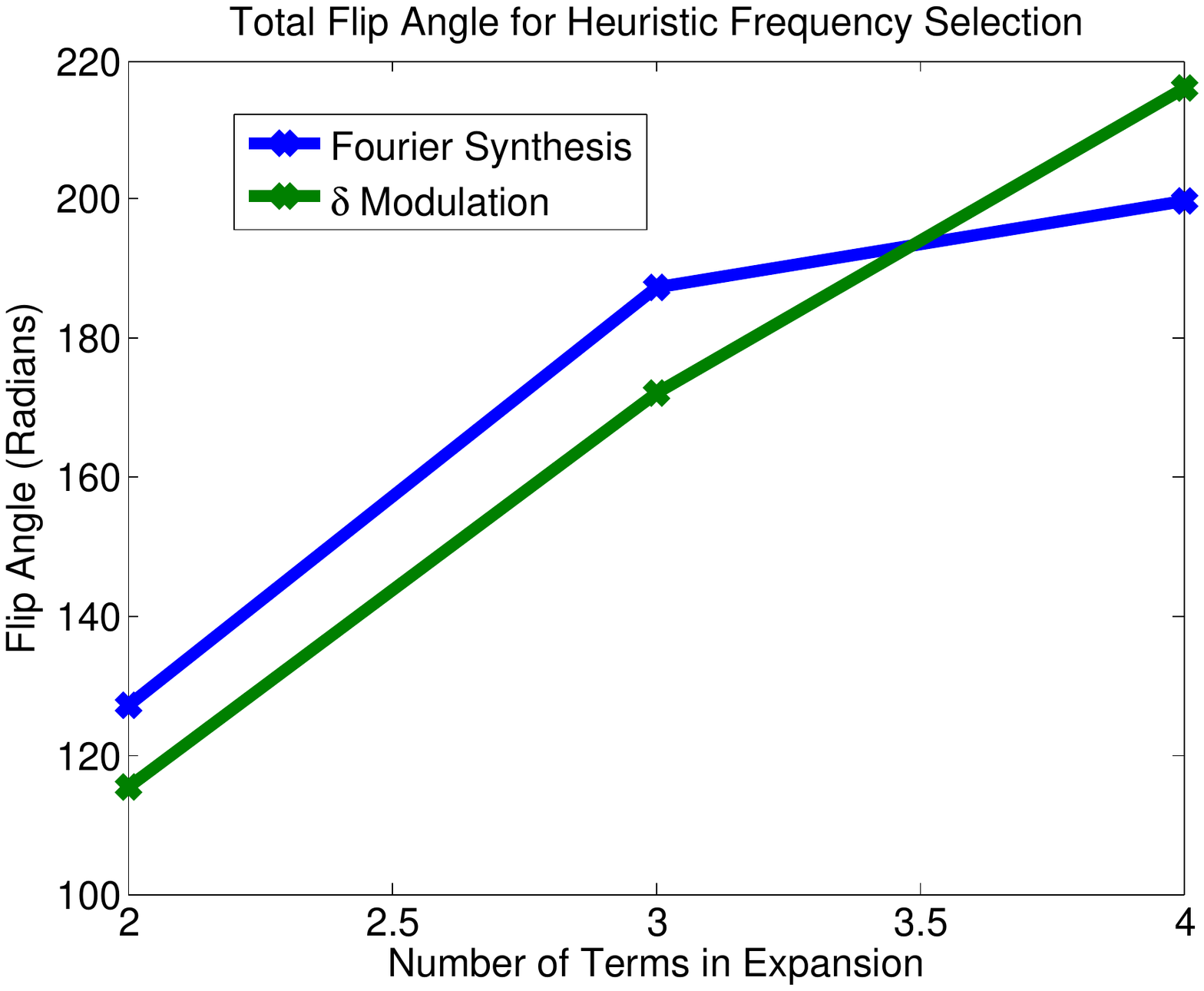}
\end{minipage}
\caption{L2 Error (Left) and Total Flip Angle (Right) for Heuristic Frequency Selection for Fourier Synthesis and $\delta$ Modulation.  Error for a Fixed Pulse Duration is Smaller for $\delta$ Modulation.}
\end{figure*}

\begin{figure*}[t!]
\begin{minipage}[b]{.95\columnwidth}
\centering
\includegraphics[width=\columnwidth,trim=30 180 60 180,clip=]{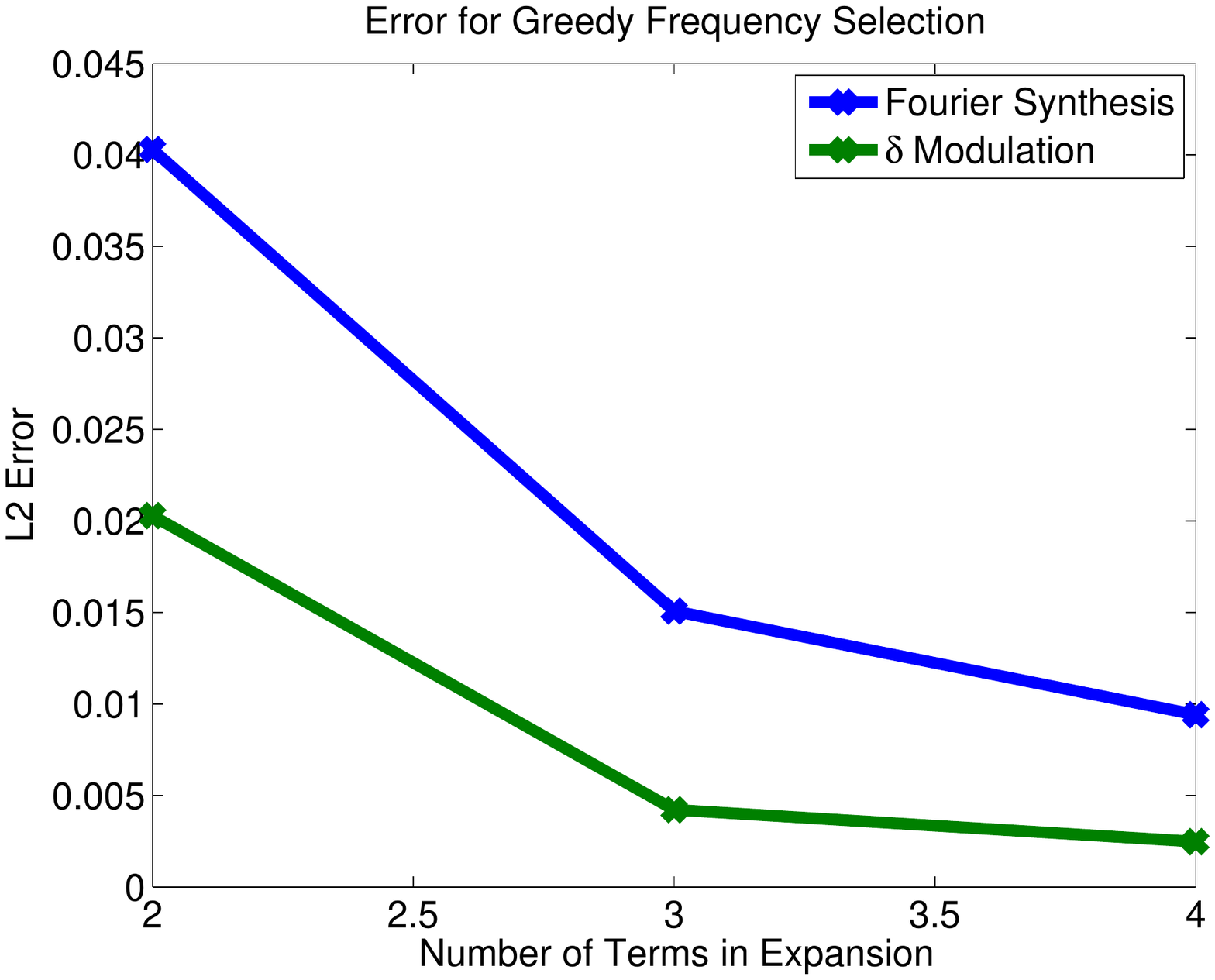}
\end{minipage} \hspace{1cm}
\begin{minipage}[b]{.95\columnwidth}
\centering
\includegraphics[width=\linewidth,trim=30 180 60 180,clip=]{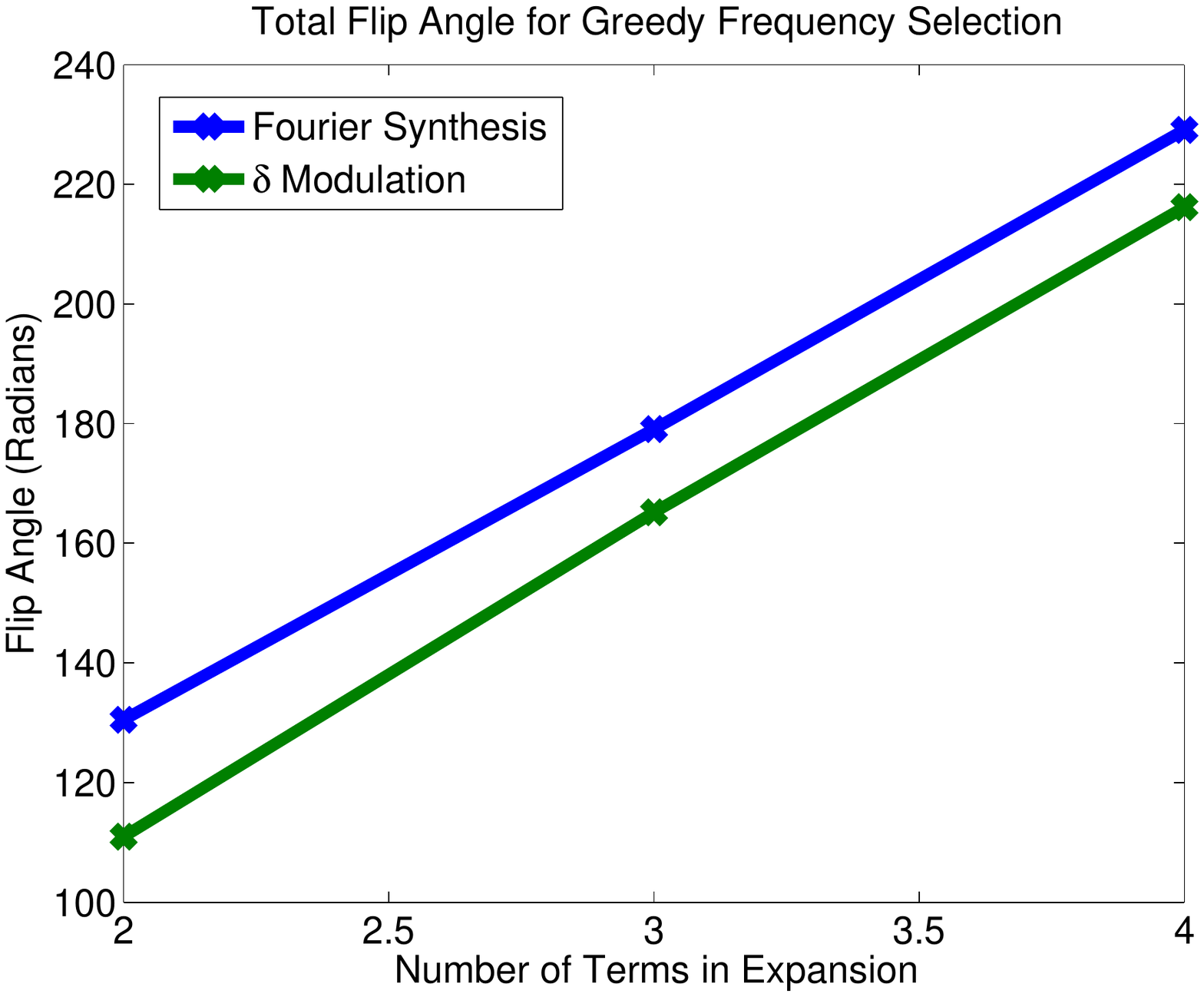}
\end{minipage}
\caption{L2 Error (Left) and Total Flip Angle (Right) for Greedy Frequency Selection for Fourier Synthesis and $\delta$ Modulation.  Both Duration and Error are Smaller for $\delta$ Modulation.}
\end{figure*}

\begin{figure*}[t!]
\begin{minipage}[b]{.95\columnwidth}
\centering
\includegraphics[width=\columnwidth,trim=40 180 60 180,clip=]{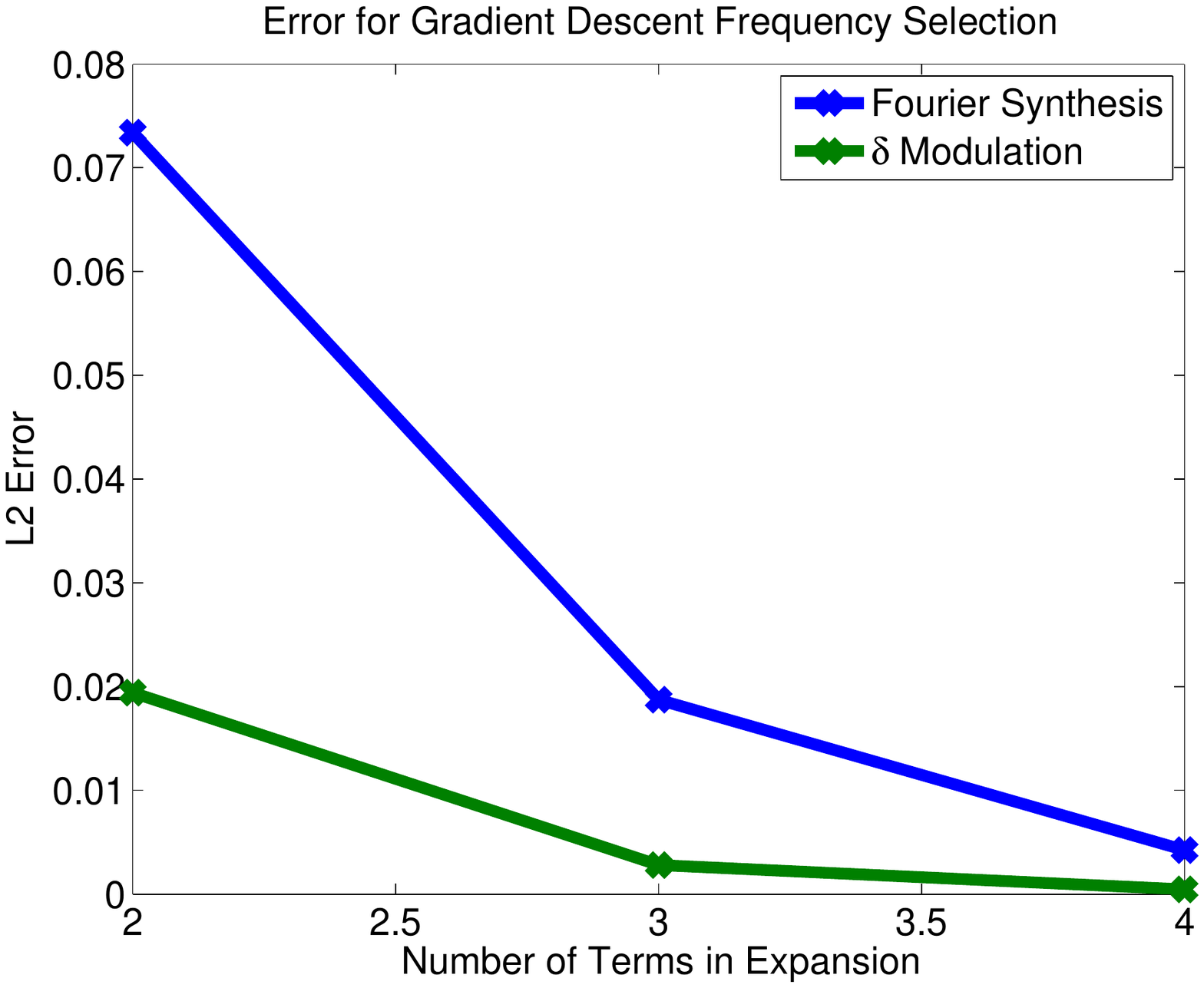}
\end{minipage} \hspace{1cm}
\begin{minipage}[b]{.95\columnwidth}
\centering
\includegraphics[width=\linewidth,trim=30 180 60 180,clip=]{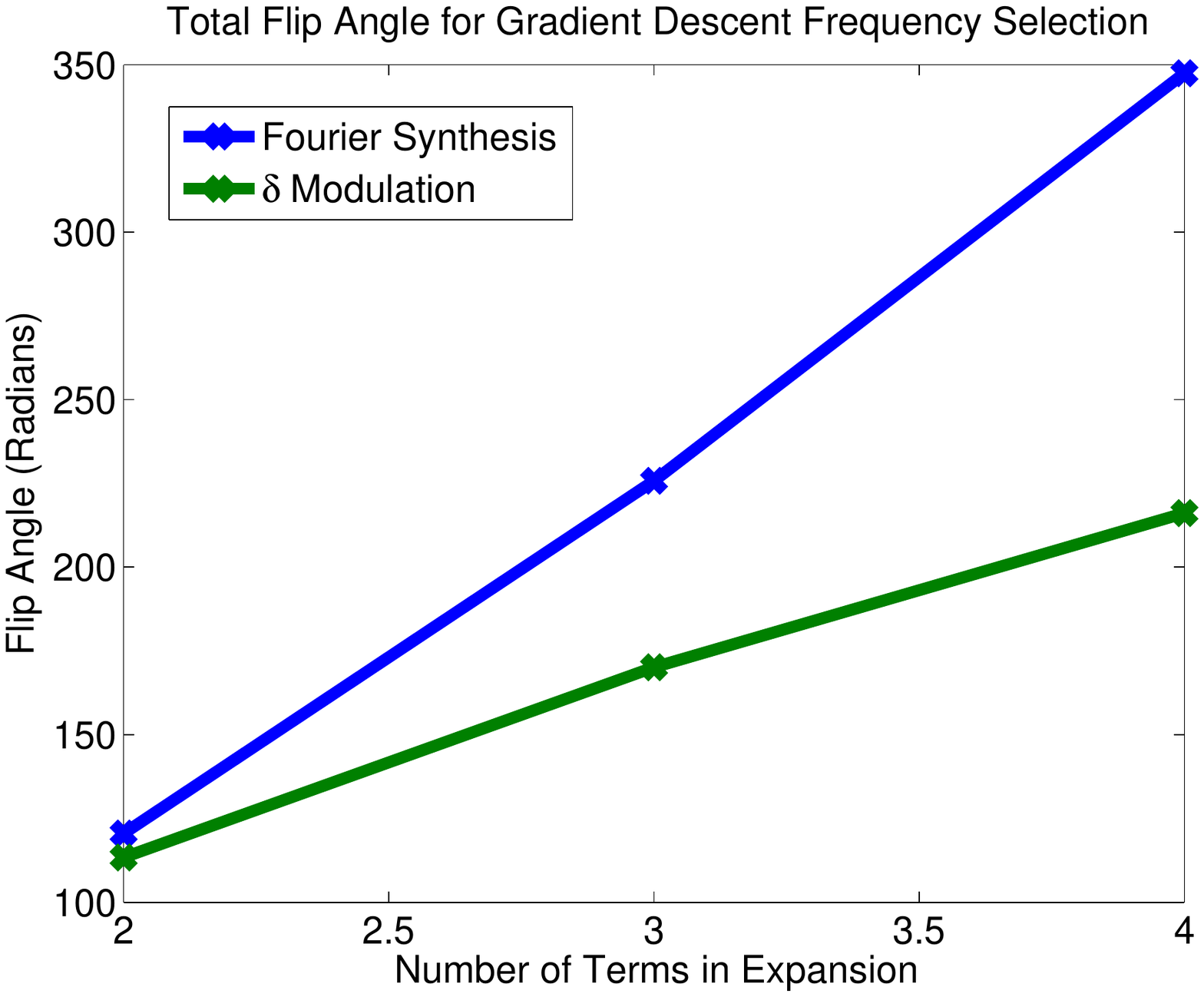}
\end{minipage}
\caption{L2 Error (Left) and Total Flip Angle (Right) for Gradient Descent Frequency Selection for Fourier Synthesis and $\delta$ Modulation.  Both Duration and Error are Smaller for $\delta$ Modulation. }
\end{figure*}

\noindent Returning to equation (\ref{eq:bloch_y}) with $\Delta t=\frac{\pi}{A}$ and
\begin{eqnarray}
\phi_1 = \begin{cases} 
0 & 0<t \leq \Delta t \\
\pi & \Delta t < t \leq 3\Delta t \\
0 & 3\Delta t < t \leq 4\Delta t
\end{cases}
\end{eqnarray}
\begin{eqnarray}
\dot{Y} = (A \ep \cos \phi_1(t)\Omega_x - \dot{\phi}_2\Omega_z)Y
\end{eqnarray}
Moving into the interaction frame 
\begin{eqnarray}
Z &\equiv & \exp(-\int \epsilon A\cos \phi_1(t)dt \  \Omega_x)Y  \\
\dot{Z} &=& -\dot{\phi}_2 (\cos \epsilon \int A \cos \phi_1(t)dt  \  \Omega_z \nonumber \\
&& \qquad +  \sin \epsilon  \int A \cos \phi_1(t)dt \  \Omega_y )Z
\end{eqnarray}
We note that under the assumptions 
\begin{eqnarray}
\int_0^T \cos \phi_1(t)dt = \int_0^T \dot{\phi}_2(t)dt = 0
\end{eqnarray}
the Z frame will agree with the Y frame which will agree with the lab frame at time T, so that it is sufficient to analyze the system in the interaction frame.  Letting $\phi_2(t)$ be a linear modulation of the form
\begin{eqnarray}
\dot{\phi}_2 = \begin{cases} 
-B & 0 \leq t < 2\Delta t  \\
B & 2\Delta t \leq t \leq 4\Delta t \\
\end{cases}
\end{eqnarray}
we can analyze the resulting rotation with the Peano-Baker Series
\begin{eqnarray}
\Phi &=& I + \int_0^{\frac{4\pi}{A}} H(t)dt + \int_0^{\frac{4\pi}{A}} H(t)\int_0^t H(\sigma_1)d\sigma_1 dt + ...  \nonumber\\
&=& I + \int_0^{\frac{4\pi}{A}} H(t)dt + o\left( \frac{B}{A} \right)^2  \nonumber \\
&=& I + 4B\int_0^{\frac{\pi}{A}} \sin(A\epsilon t)dt \Omega_y + O\left( \frac{B}{A} \right)^2 \nonumber \\
&=& I + \frac{4B}{A\epsilon} (1-\cos(\pi \epsilon)) + O\left( \frac{B}{A} \right)^2  \nonumber \\
&=& I + \frac{8B}{A} \left( 1-\delta  \right) + o(\delta)^2 + O\left( \frac{B}{A} \right)^2
\end{eqnarray}
which, to first order, has resulted in an evolution with the dispersion term reversed.  Combining with a directly accessible evolution of $\frac{8B(1+\delta)}{A}\Omega_y$ will produce a pulse that is robust to first order in $\delta$.  Figure 3 displays the trajectory in the interaction frame for the linearly modulated pulse with $\ep>1$ and provides the intuition for why the dispersion term is negated.

\subsection*{Arbitrary Modulation Schemes}

\noindent Other modulation functions are also possible.  Let $|f(t)|<B$ be such a candidate modulation, then choosing
\begin{eqnarray}
\dot{\phi}_2 = \begin{cases} 
f(t) & 0<t \leq \Delta t \\
f(2\Delta t -t) & \Delta t < t \leq 2\Delta t \\
-f(t-2\Delta t) & 2\Delta t < t \leq 3\Delta t \\
-f(4\Delta t - t) & 3\Delta t < t \leq 4\Delta t
\end{cases}
\end{eqnarray}
will produce a rotation
\begin{eqnarray}
I - 4\int_0^{\frac{\pi}{A}} f(t)\sin(A\ep t)dt\Omega_y + O\left( \frac{B}{A} \right)^2
\end{eqnarray}
which can be used to produce new dispersion dependencies and thereby robust pulses as was done in the linear case.

\section{Conclusion}

We have presented a new method for pulse design in the presence of RF-inhomogeniety that extends existing Fourier Synthesis Methods.  The method displays superior time-error properties to conventional Fourier Synthesis Methods.  This method is analytical and can be used to produce arbitrarily robust performance.       

%Recent advancements...
%
%Numerical methods exist (cite new jr shin paper?) for larmor and rf...need analytical...

\begin{widetext}

\section{Pulse Parameter Appendix}

%Heuristic FSM
Heuristic FSM, n=2
\begin{eqnarray*}
\alpha &=& [187.3,  33.8]  \\
\gamma &=& [49.3,  196.5] \\
Pulse:  && [(49.3)_0(4.5)_{90}(98.5)_{180}(4.5)_{90}(49.3)_0]^{\times 21}[(196.5)_0(4.2)_{90}(393.0)_{180}(4.2)_{90}(196.5)_0]^{\times 4}
\end{eqnarray*}

Heuristic FSM, n=3
\begin{eqnarray*}
\alpha &=& [201.1,  49.2,   7.3]  \\
\gamma &=& [49.3,  196.5,  369.0] \\
Pulse:   && [(49.3)_0(4.4)_{90}(98.5)_{180}(4.4)_{90}(49.3)_0]^{\times 23} \\
&& \qquad [(196.5)_0(4.1)_{90}(393.0)_{180}(4.1)_{90}(196.5)_0]^{\times 6}[(369.0)_0(3.6)_{90}(738.0)_{180}(3.6)_{90}(369.0)_0]^{\times 1}
\end{eqnarray*}

Heuristic FSM, n=4
\begin{eqnarray*}
\alpha &=& [175.2903,  18.3977,   -10.8059,    -5.67454]  \\
\gamma &=& [49.3,  196.5,  369.0, 546.0] \\
Pulse:  && [(49.3)_0(4.4)_{90}(98.5)_{180}(4.4)_{90}(49.3)_0]^{\times 20}[(196.5)_0(3.1)_{90}(393.0)_{180}(3.1)_{90}(196.5)_0]^{\times 3} \\
&& \qquad [(369.0)_0(-2.7)_{90}(738.0)_{180}(-2.7)_{90}(369.0)_0]^{\times 2}[(546.0)_0(-2.8)_{90}(1092.1)_{180}(-2.8)_{90}(546.0)_0]^{\times 1}
\end{eqnarray*}

%Heuristic delta
Heuristic $\delta$ Mod, n=2
\begin{eqnarray*}
\alpha &=& [105.5, 16.7]  \\
\gamma &=& [90,270] \\
Pulse:   && [(90.0)_0(180.0)_{175.6}(90.0)_0]^{\times 12}[(270.0)_0(540.0)_{175.8}(270.0)_0]^{\times 2}
\end{eqnarray*}

Heuristic $\delta$ Mod, n=3
\begin{eqnarray*}
\alpha &=& [108.3,22.4,4.3]  \\
\gamma &=& [90,270,450] \\
Pulse:   && [(90.0)_0(180.0)_{175.8}(90.0)_0]^{\times 13}[(270.0)_0(540.0)_{176.3}(270.0)_0]^{\times 3}[(450.0)_0(900.0)_{177.9}(450.0)_0]^{\times 1}
\end{eqnarray*}

Heuristic $\delta$ Mod, n=4
\begin{eqnarray*}
\alpha &=& [109.8,25.7,7.1,1.2]  \\
\gamma &=& [90,270,450,630] \\
Pulse:  && [(90.0)_0(180.0)_{175.8}(90.0)_0]^{\times 13}[(270.0)_0(540.0)_{175.7}(270.0)_0]^{\times 3} \\
&& \qquad [(450.0)_0(900.0)_{176.4}(450.0)_0]^{\times 1}[(630.0)_0(1260.0)_{179.4}(630.0)_0]^{\times 1}
\end{eqnarray*}

%Greedy FSM
Greedy FSM, n=2
\begin{eqnarray*}
\alpha &=& [191.9,35.9]  \\
\gamma &=& [49.9,192.7] \\
Pulse:  && [(49.9)_0(4.4)_{90}(99.9)_{180}(4.4)_{90}(49.9)_0]^{\times 22}[(192.7)_0(4.5)_{90}(385.4)_{180}(4.5)_{90}(192.7)_0]^{\times 4}
\end{eqnarray*}

Greedy FSM, n=3
\begin{eqnarray*}
\alpha &=& [197.4,40.9,-3.8]  \\
\gamma &=& [49.9,192.7,502.9] \\
Pulse:   && [(49.9)_0(4.5)_{90}(99.9)_{180}(4.5)_{90}(49.9)_0]^{\times 22}[(192.7)_0(4.1)_{90}(385.4)_{180}(4.1)_{90}(192.7)_0]^{\times 5} \\
&& \qquad [(502.9)_0(-1.9)_{90}(1005.8)_{180}(-1.9)_{90}(502.9)_0]^{\times 1}
\end{eqnarray*}

Greedy FSM, n=4
\begin{eqnarray*}
\alpha &=& [200.7,43.7,-5.9,-1.9]  \\
\gamma &=& [49.9,192.7,502.9,666.8] \\
Pulse:  && [(49.9)_0(4.4)_{90}(99.9)_{180}(4.4)_{90}(49.9)_0]^{\times 23}[(192.7)_0(4.4)_{90}(385.4)_{180}(4.4)_{90}(192.7)_0]^{\times 5} \\
&& \qquad [(502.9)_0(-3.0)_{90}(1005.8)_{180}(-3.0)_{90}(502.9)_0]^{\times 1}[(666.8)_0(-0.9)_{90}(1333.7)_{180}(-0.9)_{90}(666.8)_0]^{\times 1}
\end{eqnarray*}

%Greedy delta
Greedy $\delta$ Mod, n=2
\begin{eqnarray*}
\alpha &=& [105.5,16.6]  \\
\gamma &=& [86.7,259.1] \\
Pulse:   && [(86.7)_0(173.4)_{175.6}(86.7)_0]^{\times 12}[(259.1)_0(518.1)_{175.8}(259.1)_0]^{\times 2}
\end{eqnarray*}

Greedy $\delta$ Mod, n=3
\begin{eqnarray*}
\alpha &=& [108.2,22.2,4.1]  \\
\gamma &=& [86.7,259.1,427.8] \\
Pulse:   && [(86.7)_0(173.4)_{175.8}(86.7)_0]^{\times 13}[(259.1)_0(518.1)_{176.3}(259.1)_0]^{\times 3}[(427.8)_0(855.7)_{177.9}(427.8)_0]^{\times 1}
\end{eqnarray*}

Greedy $\delta$ Mod, n=4
\begin{eqnarray*}
\alpha &=& [108.5,22.9,4.6,-.3]  \\
\gamma &=& [86.7,259.1,427.8,730.2] \\
Pulse:  && [(86.7)_0(173.4)_{175.8}(86.7)_0]^{\times 13}[(259.1)_0(518.1)_{176.2}(259.1)_0]^{\times 3} \\
&& \qquad [(427.8)_0(855.7)_{177.7}(427.8)_0]^{\times 1}[(730.2)_0(1460.5)_{180.2}(730.2)_0]^{\times 1}
\end{eqnarray*}

%Gradient FSM
Gradient Descent FSM, n=2
\begin{eqnarray*}
\alpha &=& [163.4,-15.7]  \\
\gamma &=& [51.5,373.7] \\
Pulse:  && [(51.5)_0(4.3)_{90}(103.0)_{180}(4.3)_{90}(51.5)_0]^{\times 19}[(373.7)_0(-3.9)_{90}(747.4)_{180}(-3.9)_{90}(373.7)_0]^{\times 2}
\end{eqnarray*}

Gradient Descent FSM, n=3
\begin{eqnarray*}
\alpha &=& [169.6,-23.9,-10.3]  \\
\gamma &=& [52.4,379.1,550.3] \\
Pulse:   && [(52.4)_0(4.5)_{90}(104.9)_{180}(4.5)_{90}(52.4)_0]^{\times 19}[(379.1)_0(-4.0)_{90}(758.2)_{180}(-4.0)_{90}(379.1)_0]^{\times 3} \\
&& \qquad [(550.3)_0(-2.6)_{90}(1100.6)_{180}(-2.6)_{90}(550.3)_0]^{\times 2}
\end{eqnarray*}

Gradient Descent FSM, n=4
\begin{eqnarray*}
\alpha &=& [174.4,-30.6,-19.0,-5.1]  \\
\gamma &=& [53.1,381.4,554.0,727.9] \\
Pulse:  && [(53.1)_0(4.4)_{90}(106.1)_{180}(4.4)_{90}(53.1)_0]^{\times 20}[(381.4)_0(-3.8)_{90}(762.7)_{180}(-3.8)_{90}(381.4)_0]^{\times 4} \\ 
&& \qquad [(554.0)_0(-3.2)_{90}(1108.0)_{180}(-3.2)_{90}(554.0)_0]^{\times 3}[(727.9)_0(-2.6)_{90}(1455.8)_{180}(-2.6)_{90}(727.9)_0]^{\times 1}
\end{eqnarray*}

%Gradient delta
Gradient Descent $\delta$ Mod, n=2
\begin{eqnarray*}
\alpha &=& [105.5,16.6]  \\
\gamma &=& [88.6,265.1] \\
Pulse:   && [(88.6)_0(177.1)_{175.6}(88.6)_0]^{\times 12}[(265.1)_0(530.1)_{175.9}(265.1)_0]^{\times 2}
\end{eqnarray*}

Gradient Descent $\delta$ Mod, n=3
\begin{eqnarray*}
\alpha &=& [108.3,22.4,4.3]  \\
\gamma &=& [89.1,267.0,444.5] \\
Pulse:   && [(89.1)_0(178.1)_{175.8}(89.1)_0]^{\times 13}[(267.0)_0(534.1)_{176.3}(267.0)_0]^{\times 3}[(444.5)_0(889.0)_{177.9}(444.5)_0]^{\times 1}
\end{eqnarray*}

Gradient Descent $\delta$ Mod, n=4
\begin{eqnarray*}
\alpha &=& [109.8,25.7,7.1,1.2]  \\
\gamma &=& [90.0,270.0,450.0,630] \\
Pulse:  && [(90.0)_0(180.0)_{175.8}(90.0)_0]^{\times 13}[(270.0)_0(540.0)_{175.7}(270.0)_0]^{\times 3} \\
&& \qquad [(450.0)_0(900.0)_{176.4}(450.0)_0]^{\times 1}[(630.0)_0(1260.0)_{179.4}(630.0)_0]^{\times 1}
\end{eqnarray*}

\end{widetext}

\end{document}